\documentclass[twocolumn,aps,prb,superscriptaddress,citeautoscript]{revtex4}

\usepackage{graphicx}
\usepackage{dcolumn}
\usepackage{bm}
\usepackage{amsmath}
\usepackage[T1]{fontenc}
\usepackage{sidecap}
\usepackage{float}
\usepackage{booktabs}
\usepackage{hyperref}
\hypersetup{
    colorlinks,
    linkcolor={black},
    citecolor={blue},
    urlcolor={blue}
}

\begin{document}

\preprint{}

\author{Jakub Pawlak}
 \email{jakub.pawlak@agh.edu.pl}
\affiliation{AGH University of Science and Technology, Faculty of Physics and Applied Computer Science, Al. Mickiewicza 30, 30-059 Krak\'{o}w, Poland}
\affiliation{AGH University of Science and Technology, Academic Centre for Materials and Nanotechnology, Al. Mickiewicza 30, 30-059 Krak\'{o}w, Poland}
\author{Witold Skowro\'{n}ski}
  \email{skowron@agh.edu.pl}
\affiliation{AGH University of Science and Technology, Institute of Electronics, Al. Mickiewicza 30, 30-059 Krak\'{o}w, Poland}
\affiliation{CIC nanoGUNE BRTA, 20018 Donostia-San Sebastian, Basque Country, Spain}
\author{Piotr Ku\'{s}wik}
\affiliation{Institute of Molecular Physics, Polish Academy of Sciences, ul. Smoluchowskiego 17, 60-179 Poznań, Poland}
\author{F\'{e}lix Casanova}
\affiliation{CIC nanoGUNE BRTA, 20018 Donostia-San Sebastian, Basque Country, Spain}
\author{Marek Przybylski}
\affiliation{AGH University of Science and Technology, Faculty of Physics and Applied Computer Science, Al. Mickiewicza 30, 30-059 Krak\'{o}w, Poland}
\affiliation{AGH University of Science and Technology, Academic Centre for Materials and Nanotechnology, Al. Mickiewicza 30, 30-059 Krak\'{o}w, Poland}
\title{Spin Hall Induced Magnetization Dynamics in Multiferroic Tunnel Junction}
\begin{abstract}
The combination of spin-orbit coupling driven effects and multiferroic tunneling properties was explored experimentally in thin Pt/Co/BTO/LSMO multilayers. The presence of a Pt heavy metal allows for the spin current-induced magnetization precession of Co upon radio-frequency charge current injection. The utilization of a BTO ferroelectric tunnel barrier separating the Co and LSMO ferromagnetic electrodes gives rise to both tunneling- magnetoresistance and electroresistance. Using the spin-orbit torque ferromagnetic resonance, the maganetization  dynamics of the Co/Pt bilayers was studied at room temperature. Unexpectedly the magnetization dynamics study in the same geometry performed at low temperature reveals the existence of both Co and LSMO resonance peaks indicating efficient spin current generation both using the spin Hall effect in Pt and spin pumping in LSMO that tunnel via the BTO barrier. 

\end{abstract}

\maketitle

\section{Introduction}

The utilization of magnetic elements and spin current in spintronics devices enabled the development of magnetic random access memories \cite{tehrani1999} and magnetic field sensors used in storage devices \cite{Dieny-2020}. The use of spin as an object representing information, due to its physical properties, makes it theoretically possible to obtain devices that switch faster, with lower energy consumption, reduce Joule heating and are non-volatile at the same time.
One of the original implementation limitations of the achievements in this field was the necessity to use an external magnetic field to control the device. With ongoing development, it has become possible to replace the field with a polarized current flowing through the junction (STT) \cite{SLONCZEWSKI1996L1}, and then with the current flowing only through the upper electrode (SOT) \cite{manchon2009}. While the method of reading information based on the inverse spin Hall effect seems to be sufficient, the SOT (using the spin Hall effect) requires a very high current density. 
The concept to further decrease a possible energy consumption of spintronic elements is the so-called MESO \cite{Manipatruni2019}, which combines the magnetoelectric writing \cite{Chu2008, Heron2014} of the magnetization state and the inverse-spin Hall effect \cite{Pham2020} or the inverse Rashba–Edelstein effect \cite{Manchon2015} reading of the stored information.
 The implementation of this idea makes use of a ferroelectric material, also used in FeRAM memories \cite{MIKOLAJICK2001947}. The complex relationships between spin and ferroic effects are at the initial stage of research and still require fundamental, theoretical as well as application-technological research. Much of the contemporary work focuses on devices based on oxide materials exhibiting ferroelectricity, high spin polarization, phase transitions, etc., in combination with materials with high spin-orbit coupling, such as a heavy metal and ferromagnetic layer \cite{cho_large_2015}.
 An alternative approach to realize this concept is the multiferroic tunnel junction (MFTJ) \cite{Gajek2007} with an addition of a spin-orbit coupling material. 
 Very interesting phenomena with a high application potential could be observed. As it has recently been shown, an interfacial spin-orbit torque (SOT) can be tuned with the ferroelectric polarization of the PZT underlayer \cite{Fang2020}.

In this work, a MFTJ with a Co-based top electrode covered with Pt, which is characterized by a moderate spin-orbit coupling, is presented \cite{Liu2011}.
The bottom part of the device is composed of a BaTiO$_3$ (BTO) ferroelectric tunnel barrier grown on an epitaxial ferromagnetic, half-metallic oxide La$_{2/3}$Sr$_{1/3}$MnO$_3$ (LSMO) \cite{huijben2008}. 
The MFTJ we reported in a previous work \cite{pawlak_2022}, based on the same oxide materials, showed multiferroicity at room temperature, for the first time for an oxide electrode.
The combination of these materials gives the opportunity to study spin effects and multiferroic properties as well as the interplay between them.
The SOT is determined by the angular dependent harmonic Hall voltage measurements. Both the MFTJ characteristics, including tunneling magnetoresistance (TMR) and tunneling electroresistance (TER) as well as the spin-Hall magnetoresistance (SMR) are present. The magnetization dynamics was determined using the spin-orbit torque ferromagnetic resonance (SOT-FMR) at different temperatures, which enables the extraction of the magnetic properties of the ferromagnetic electrodes. The characteristic peak of Co is present, with an additional second peak, measured at lower temperatures, which is associated with LSMO. This is unexpected, as the SOT-induced dynamics is measured with an in-plane current flow configuration. Thus we were able to design and fabricate a device that shows the TER and TMR as well as SOT-induced magentization dynamics of both ferromagnetic electrodes.

\begin{figure}[t]
\centering
\includegraphics[width=\columnwidth]{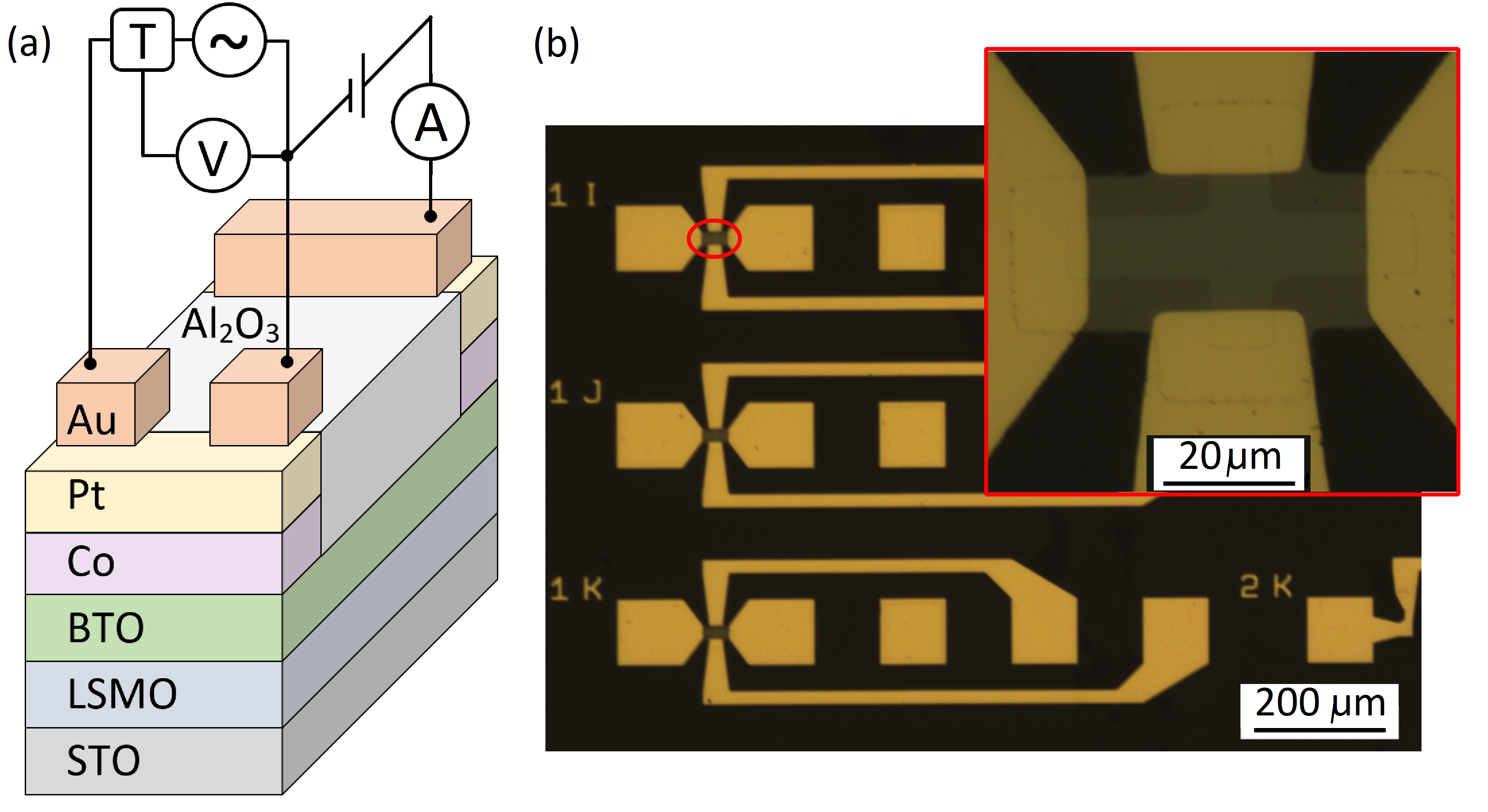}
\caption{Pt/Co/BTO/LSMO/STO multilayer system with the measurements setup sketched (a). The SMR/AMR and the spin Hall measurement require an in-plane current flow, whereas the tunneling properties are measured with the current tunneling via the BTO barrier. The SOT-FMR was performed using a microwave signal generator connected to the Co/Pt bilayer using the capacitor terminal bias-T, while the mixing voltage was measured using the inductance terminal. The micrograph of the fabricated devices is presented in (b), with an  inset presenting a close-up on the Hall cross.}
\label{fig:structure}
\end{figure}

\section{Experiment}
Two series of Pt/Co/BTO/LSMO samples forming a multiferroic tunnel junctions were prepared. One with a Co wedge (from 1 to 4 nm) with a thickness of Pt kept constant $t_\mathrm{Pt}$ = 5 nm and the other one with a Pt wedge (from 0 to 10 nm) with a constant thickness of Co $t_\mathrm{Co}$ = 1 nm. The thickness of the BTO and LSMO were 3.5 and 3 nm, respectively. This enabled us to test the properties for different ferromagnet and heavy metal thicknesses, while maintaining the properties of the remaining layers. The samples were patterned in a four-step optical lithography process in the following way. First, the LSMO and BTO oxide layers were deposited at a temperature of $T$ =  740 $^{\circ}$C and an oxygen pressure of $P$ = 200 mTorr \cite{pawlak2018,pawlak2020} on TiO-terminated STO substrates \cite{pawlak2021} using the pulsed laser deposition (PLD) technique in a Necocera Pioneer system equipped with a Coherent Compex 110F laser. Then, the first lithography soft-mask was prepared and the sample was covered with a 30-nm-thick Al$_2$O$_3$ insulating layer using the magnetron sputtering technique. After the lift-off process, the second lithography step was prepared with rectangle and cross-shape devices following the Co and Pt wedges deposition by magnetron sputtering. Thw wedges were realized thanks to the moving shutter during the deposition. 
Finally, after the third-lithography step, Ti(5 nm)/Au(50 nm) contact electrodes were deposited following the lift-off process. 
The schematics of the multilayer structure and the micrographs of the  fabricated devices are presented in Fig. 1. The Hall-crosses are 10 $\mu$m wide and 50 $\mu$m long. Additional rectangles of 10 $\times$ 10 $\mu$m and 5 $\times$ 5 $\mu$m
were also prepared for the TMR and TER measurements. The spin Hall efficiency of the Co/Pt bilayer was determined in the Hall-cross geometry, with a current flowing in the Co/Pt plane using the angular dependent harmonic Hall voltage measurement \cite{avcii_2014} in the rotating probe-station co-developed with Measline Ltd. The spin-orbit torque ferromagnetic resonance (SOT-FMR) \cite{Tulapurkar2005} was determined also in the current in-plane configuration, both using the two-point \cite{Liu2011} and four-point (transverse) \cite{lazarski2021} geometry. Tunneling via the BTO barrier was measured in the two-point geometry between the top Co/Pt bilayer and the bottom (unpatterned) LSMO electrode. The temperature dependent measurements were performed in a cryogenic free probe station in the temperature range between 10 and 300 K. Additional FMR studies were performed on the LSMO thin film using a vector-network analyzer and coplanar waveguide mounted in the cryostat.

\section{Results and discussion}
    
    \subsection{Magneto-transport characterisation of Co/Pt bilayer}

\begin{figure}[t]
\centering
\includegraphics[width=\columnwidth]{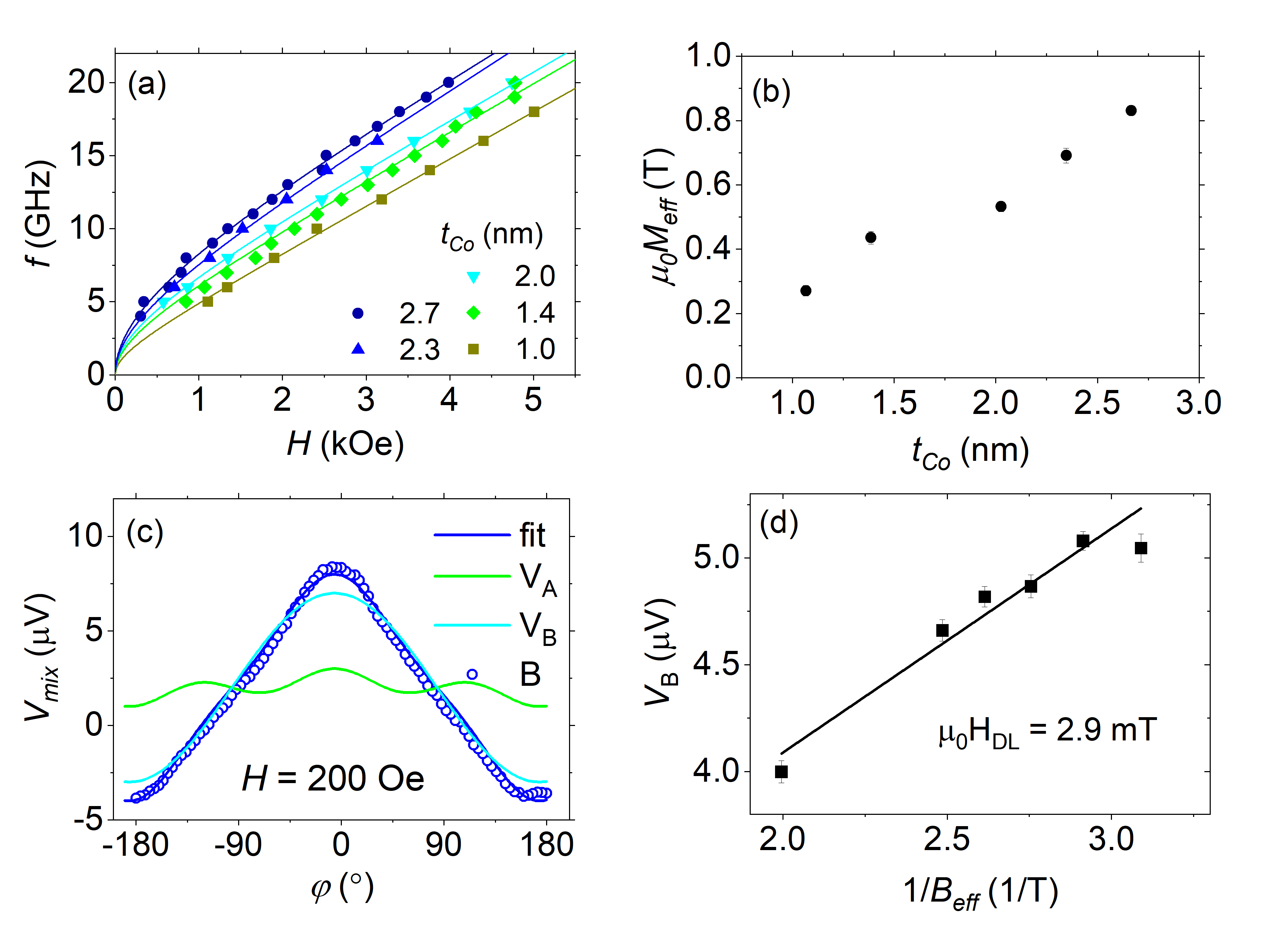}
\caption{Magneto-transport characterization of Co-Pt top bilayer. Resonance frequency $f$ vs. external magnetic field $H$ (a) and the extracted effective magnetization of Co as a result of fitting theoretical $f$ vs. $H$ dependence using a Kittel formula (b). An example of angular dependence of the second-harmonic Hall voltage for $H$ = 200 Oe together with a fit representing field-like term $V_A$ and damping-like term $V_B$ (c). The amplitude of the damping-like effective field voltage as a function of the inverse effective magnetic field induction enables damping-like effective field calculation (d). }
\label{fig:co-pt}
\end{figure}

    The multilayer system is composed of two parts: the top Co/Pt bilayer, which exhibits spin-orbit coupling, and Co/BTO/LSMO MFTJ, with tunneling magnetoresistance and electroresistance present. The resistivities of Co and Pt were extracted from their thickness dependence. The magnetic properties of Co were determined based on the analysis of the magnetization dynamics obtained via SOT-FMR \cite{Liu_luqiao2011}. The resonance frequency vs. magnetic field dependencies were fitted using a Kittel formula and yielded the effective magnetization of Co of different thickness - Fig. \ref{fig:co-pt}(a-b) 
    As expected, the effective magnetization of Co decreases with the layer thickness from around $M_\mathrm{eff}$ = 0.55 T for $t_\mathrm{Co}$ = 2.8 nm down to $M_\mathrm{eff}$ = 0.15 T for $t_\mathrm{Co}$ = 1 nm. Unlike with the sample covered with MgO \cite{lazarski_2019}, effective perpendicular magnetic anisotropy in Co is absent (see Supplemental Information for the detailed analysis). 
    The example of a second harmonic Hall voltage measurement ($V_{\mathrm{2\omega}}$) \cite{hayashi2014} as a function of the in-plane magnetic field angle ($\phi$) in the Co(3)/Pt(5) bilayer presented in Fig. \ref{fig:co-pt}(c) exhibits a characteristic dependence that was fitted with Eq:
    \begin{equation}
        V_{\mathrm{2\omega}} = -V_B\cos\varphi \cos2\varphi - V_A\cos\varphi
    \label{eq:Vharm}
    \end{equation}
    where $V_\mathrm{B}$ and $V_\mathrm{A}$ are the amplitudes of the terms proportional to the planar and anomalous Hall effects (AHE). The amplitude of the AHE was determined from a separate measurement performed in the perpendicular orientation ($H$ along the $z$ axis). Based on the linear fit of the $V_\mathrm{B}$ vs. the reciprocal of the effective field, the amplitude of the damping-like effective field was calculated - Fig. \ref{fig:co-pt}(d). The spin Hall efficiency $\xi_\mathrm{DL}$ was calculated using the formula \cite{Pai2015}:
    \begin{equation}
        \xi_\mathrm{DL} = \cfrac{\mu_\mathrm{0} M_\mathrm{S} t_\mathrm{Co} H_\mathrm{DL}}{j \hbar/2e}
    \label{eq:ksi}
    \end{equation}
    The value $\xi_\mathrm{DL}$ = 0.14 was obtained with a negligible contribution of the field-like term, similar to our previous studies \cite{skowronski2021}.

    Apart from the SOT, the top Co/Pt bilayer also exhibits anisotropic magentoresistance (AMR) and spin Hall magnetoresistance (SMR). In order to separate them, the angular dependence of the resistance in the $\alpha$, $\beta$ and $\gamma$ planes was measured \cite{Althammer2013} and analysed (see Supplemental Information).  
    
    \subsection{Tunneling magnetoresitance and electroresistance}
    
\begin{figure}[t]
\centering
\includegraphics[width=\columnwidth]{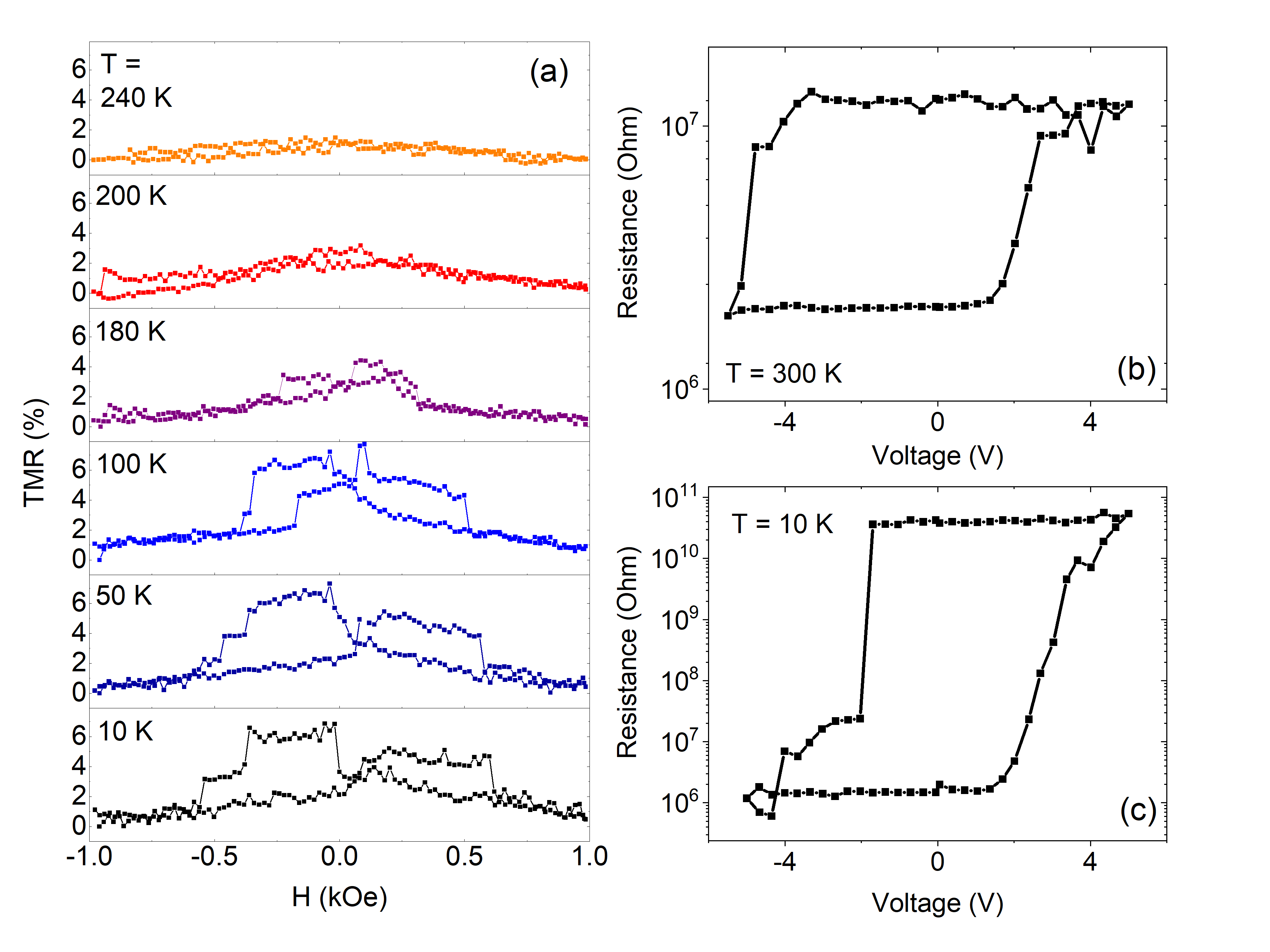}
\caption{The tunneling properties of the MFTJ. The TMR is measured at temperatures below 200 K. At higher temperatures, only the remaining CMR originating from the bottom LSMO electrode is present (a). The TER ratio increases from one order of magnitude at $T$ = 300 K (b) up to four orders of magnitude at $T$ = 10 K (c). }
\label{fig:mftj}
\end{figure}

    Next, we proceed to the analysis of the tunneling properties of the fabricated devices. Apart from the AMR and SMR, the MFTJ is also characterized by TMR and TER, while in the bottom LSMO electrode collossal magnetoresistance (CMR) is present. In order to separate between these effects, different measurement configurations and temperatures were employed. The AMR/SMR was measured using only the top electrodes, while the CMR was determined with the current flowing only through the bottom electrode. The TMR and TER were measured with the current tunneling through the BTO ferroelectric barrier. Unlike in our previous work on Fe/BTO/LSMO MFTJ \cite{pawlak_2022}, the TMR is absent at room temperature, similar to other studies on Co/BTO structures \cite{garcia2014, mao2014, valencia2011}. The pseudo-spin valve behaviour in the $R$ vs. $H$ dependencies starts to be visible at $T$ < 200-260 K (depending on the devices) - Fig. \ref{fig:mftj}(a). However, the use of a slightly thicker BTO barrier, with respect to our previous studies, enabled us to increase the TER ratio from one order of magnitude at room temperature up to four orders of magnitude at T = 10 K - Fig. \ref{fig:mftj}(b-c). 
    We were able to measure the TMR only in a low resistance state. In the high-resistance polarity state of the BTO the tunneling current is reduced significantly and prevents low-noise measurements \cite{tornos2019}.

\begin{figure}[t]
\centering
\includegraphics[width=\columnwidth]{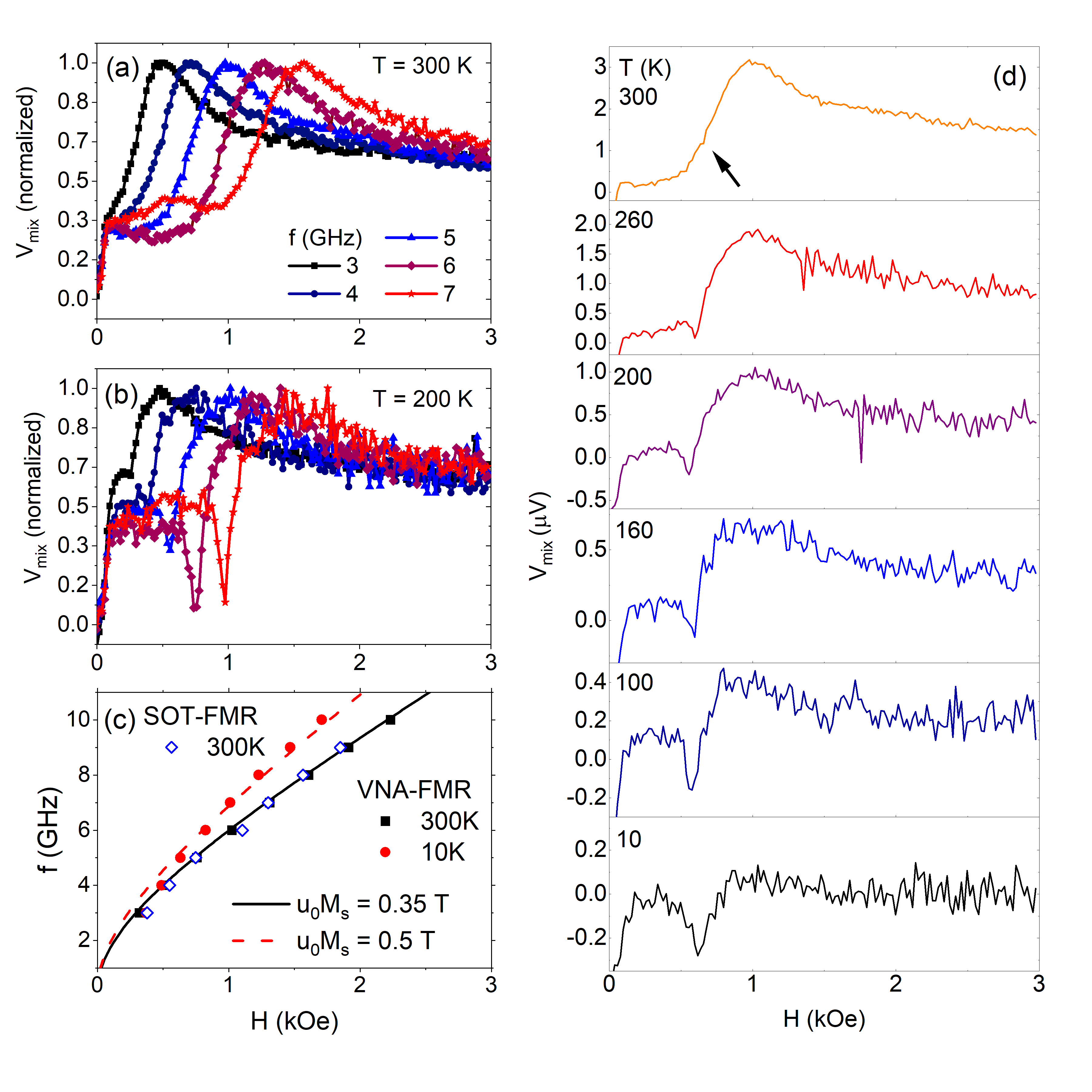}
\caption{The FMR measurement summary. At room temperature, only the broad peak originating from Co is visible (a). The measurements of the SOT-FMR spectra at $T$ = 200 K performed at different excitation frequencies reveal two resonance fields (b). The evolution of the resonance frequency vs. field for room temperatures and 10 K measured using the inductive (VNA-FMR) and electrical (SOT-FMR) methods. The lines correspond to the effective magnetization of $\mu_0 M_\mathrm{eff}$ = 0.35 $T$ (solid) and $\mu_0 M_\mathrm{eff}$ = 0.5 at (dashed) (c). The FMR measurements at constant $f$ = 5 GHz performed with decreasing temperature reveal the emergence of the second peak (d). An arrow indicates a possible position of the low-intensity, narrow linewidth peak at room temperature.}

\label{fig:dynamics}
\end{figure}

\subsection{Magnetization dynamics}
Finally, we move on to the discussion of the temperature dependence of the SOT-FMR. The room temperature measurements revealed only the peak from the Co layer, as expected in the two-point in-plane measurement configuration. Both the evolution of the resonance frequency and the linewidth as a function of the magnetic field agree well with the magnetization dynamics of Co - Fig. \ref{fig:dynamics}(a). 
Surprisingly, the SOT-FMR measurements performed below room temperature revealed an interesting feature of the MFTJ. With decreasing temperature the Co-peak intensity decreases and an additional, sharper peak shows up - Fig. \ref{fig:dynamics}(b). It starts to be noticeable below room temperature and dominates the spectrum at temperatures below 100 K. Both the linewidth and the peak's resonance magnetic field dependence on the frequency suggest that the signal comes from the LSMO bottom electrode, which is not obvious from the excitation of the Co/Pt bilayer alone. The most possible scenario of the origin of the LSMO peak is the following. The RF Oersted field associated with a current flow induces the precession of LSMO magnetization, which due to the spin-pumping effect \cite{mosendz2010}, induces the spin-current flow in the vertical direction through the BTO barrier. Such a spin current generates a DC voltage in the Co/Pt bilayer as a result of the inverse spin Hall effect (ISHE)\cite{saitoh2006}. Spin pumping across an insulator has already been reported in an Fe/MgO/Pt trilayer system and was explained by the combination of the rectification and the ISHE, which is possible due to the exchange interaction manifested by the spin mixing conductance \cite{mihalceanu2017}. This DC voltage is added to the mixing voltage present due to the conventional SOT-FMR measurement and, as a result, two resonance peaks can be visible in the field-spectrum. 
We note that the Curie temperature of LSMO is above room temperature, around T = 350 K. Moreover, close to room temperature the effective magnetization of LSMO and Co are similar. Therefore, it is also possible that two resonance peaks converge into a single one at T = 300 K. The magnetization dynamics of the unpatterned LSMO measured in the same sample using the VNA-FMR shows a clear signal at room temperature - Fig. \ref{fig:dynamics}(c). By carefully analysing the SOT-FMR spectra at room temperature, one can also see a reminiscence of the LSMO peak, which is indicated by an arrow in \ref{fig:dynamics}(d). Altogether, we conclude that in the MFTJ there is an efficient spin current tunneling between Co and LSMO via BTO. Nevertheless, we were not able to observe any dependence of the LSMO or Co magnetization dynamics on the BTO polarization, unlike in the recent work on GeTe \cite{Varotto2021}.

\section{Summary}

In summary, the magnetotransport properties of a Pt/Co/BTO/LSMO multiferroic tunnel junction were studied. The coexistence of anisotropic, colossal, spin-Hall and tunneling magnetoresistance was found, which were separated by different measurement configurations, symmetries and temperatures. The Pt/Co bilayer is characterized by a moderate spin-Hall angle, which enables magnetization dynamics excitation using the spin-orbit torque. Interestingly, at low temperatures an additional resonance is present, which most likely originates from the combination of the spin pumping in LSMO and the inverse-spin Hall effect in Pt/Co. Furthermore, the tunneling electroresistance increases with decreasing temperature reaching more than four orders of magnitude at $T$ = 10K. A high TER state prevents the measurement of the TMR and magnetization dynamics at lower temperatures. 

\section*{Acknowledgments}
We would like to thank Krzysztof Grochot for his help with data analysis and Stanisław Łazarski for the technical expertise. The research was carried out thanks to the financial support of the National Science Centre, Poland (grant No. 2017/27/N/ST5/01635) and the program “Excellence initiative research university” for the AGH University of Science and Technology. J.P. was partly supported by the EU Project POWR.03.02.00-00-I004/16. W.S. acknowledges grant No. 2021/40/Q/ST5/00209 from the National Science Centre, Poland and financial support from Polish National Agency for Academic Exchange (PPN/BEK/2020/1/00118/DEC/1). F.C. acknowledges funding by the Spanish MICINN (Project No. PID 2021-1225110B-100) and Maria de Maeztu Units of Excellence
Programme No. CEX2020-001038-M.


\bibliography{bibliography}

\end{document}